\documentclass[a4paper,journal,twocolumn,twoside,10pt]{IEEEtran}
\usepackage{ifpdf,epsfig,amsfonts,subfigure,color,amsbsy}
\usepackage{graphicx,cite,amssymb,amsmath,amsthm}
\usepackage{color}
\usepackage{rotating}
\usepackage{tikz}
\usepackage{makecell}
\usepackage{mathrsfs}
\usepackage[nolist]{acronym}
\usepackage{color}
\usepackage{booktabs}
\usepackage{multirow}
\usepackage{pgfplots,relsize}
\usepackage{cuted}

\newtheorem{definition}{Definition}

\newcommand{\sw}{\mathrm{sw}}

\begin{document}

\begin{acronym}
\acro{ACK}{acknowledgement}%
\acro{ACRDA}{asynchronous contention resolution diversity ALOHA}%
\acro{AIS}{automatic identification system}
\acro{AL}{application layer}%
\acro{AWGN}{additive white gaussian noise}%
\acro{BER}{bit error rate}%
\acro{BP}{birthday protocol}
\acro{CDF}{cumulative distribution function}%
\acro{CDMA}{code divison multiple access}%
\acro{CRA-CC}{CRA-Convolutional Code}%
\acro{CRA-SH}{CRA-Shannon Bound}%
\acro{CRA}{contention resolution ALOHA}%
\acro{CRBP}{collision resolution birthday protocol}
\acro{CRC}{cyclic redundancy check}%
\acro{CRDSA}{contention resolution diversity slotted ALOHA}%
\acro{CRDSA++}{contention resolution diversity slotted ALOHA++}%
\acro{CSA}{coded slotted ALOHA}
\acro{CSI}{channel state information}
\acro{DAMA}{demand assigned multiple access}%
\acro{DSA}{diversity slotted ALOHA}%
\acro{DVB}{digital video broadcasting}%
\acro{DVB-RCS+M}{digital video broadcasting - return channel via satellite mobility}%
\acro{DVB-RCS}{digital video broadcasting - return channel via satellite}%
\acro{DVB-S2}{digital video broadcasting - second generation}%
\acro{DVB-S2/RCS}{digital video broadcasting - second generation / return channel via satellite}%
\acro{ECRA}{enhanced contention resolution ALOHA}%
\acro{EGC}{equal gain combining}%
\acro{ESA}{European Space Agency}%
\acro{ETSI}{European Telecommunications Standards Institute}%
\acro{ETSI-BSM}{European Telecommunications Standards Institute - Braodband Satellite Multimedia}%
\acro{FDM}{frequency division multiplexing}%
\acro{FEC}{forward error correction}%
\acro{GEO}{geostationary orbit}%
\acro{GMSK}{gaussian minimum-shift keying}%
\acro{HF}{high frequency}%
\acro{IC}{interference cancellation}%
\acro{IRSA}{Irregular Repetition slotted ALOHA} %
\acro{ISO/OSI}{International Organization for Standardization/Open Systems Interconnection}%

\acro{LDPC}{low density parity check codes}%
\acro{LEO}{low earth orbit}%
\acro{LRT}{likelihood ratio test}%
\acro{LT}{Luby transform}%
\acro{M2M}{machine-to-machine}%
\acro{MAC}{medium access control}%
\acro{MAP}{maximum a-posteriori probability}%
\acro{MF}{matched filter}%
\acro{MF-TDMA}{multifrequency time division multiple access}%
\acro{ML}{maximum-likelihood}%
\acro{MMSE}{minimum mean squared error}%
\acro{MRC}{maximal-ratio combining}%
\acro{MUD}{multiuser detection}%
\acro{ND}{neighbor discovery} %

\acro{PDF}{probability density function}%
\acro{PLR}{packet loss rate}%
\acro{PMF}{probability mass function}%

\acro{RA}{random access}%
\acro{RCB}{random coding bound}%
\acro{RCS}{return channel via satellite}%
\acro{RCS+M}{return channel via satellite plus mobility}%
\acro{RF}{radio frequency}%
\acro{ROC}{receiver operating characteristics}%
\acro{RS}{Reed Solomon}%
\acro{SA}{slotted ALOHA}%
\acro{SB}{Shannon bound}%
\acro{SC}{selection combining}
\acro{S2}{second generation}%
\acro{SIC}{successive interference cancellation}%
\acro{SINR}{signal-to-interference and noise ratio} %
\acro{SN}{slot node}%
\acro{SNIR}{signal-to-noise and interference ratio}%
\acro{SNR}{signal-to-noise ratio}%
\acro{SOTDMA}{self-organized time division multiple access}
\acro{TDMA}{time division multiple access}%
\acro{UN}{user node}%
\acro{VHF}{very high frequency}%
\acro{VF}{virtual frame}


\end{acronym}

\title{Detection and Combining Techniques for Asynchronous Random Access with Time Diversity}


\author{Federico Clazzer, Francisco L\'azaro, Gianluigi Liva and Mario Marchese
\thanks{Federico Clazzer, Francisco L\'azaro and Gianluigi Liva are with the Institute of Communications and Navigation of the Deutsches Zentrum f\"{u}r Luft- und Raumfahrt (DLR), D-82234 Wessling, Germany (e-mail: \{federico.clazzer, francisco.lazaroblasco, gianluigi.liva\}@dlr.de).}
\thanks{Mario Marchese is with the Department of Electrical, Electronic, and Telecommunications Engineering, and Naval Architecture (DITEN), Via Opera Pia 13, 16145 Genova, Italy (e-mail: mario.marchese@unige.it).}}

\maketitle
\thispagestyle{empty} \pagestyle{empty}

\begin{abstract}
Asynchronous \ac{RA} protocols are particularly attractive for their simplicity and avoidance of tight synchronization requirements. Recent enhancements have shown that the use of \ac{SIC} can largely boost the performance of these schemes. A further step forward in the performance can be attained when diversity combining techniques are applied. In order to enable combining, the detection and association of the packets to their transmitters has to be done prior to decoding. We present a solution to this problem, that articulates into two phases. Non-coherent soft-correlation as well as interference-aware soft-correlation are used for packet detection. We evaluate the detection capabilities of both solutions via numerical simulations. We also evaluate numerically the spectral efficiency achieved by the proposed approach, highlighting its benefits.
\end{abstract}

\section{Introduction}


\acresetall

Sharing efficiently the resources among users that are required to access a common medium is of utmost importance in today's systems where bandwidth is scarce. Random access (RA) \acused{RA}was proposed first \cite{Abramson1970,Roberts1975} to allow users to share a common medium without coordination. Recent advances in \acs{RA} show that high efficiency can be achieved \cite{Casini2007, Liva2011, Paolini_2015_TIT,Narayanan_2012}. In these solutions, the transmitters send multiple copies of their packets (called replicas). Each replica contains information about the position of all its copies within a time slotted frame. At the receiver side, via \ac{SIC}, potential collisions are resolved taking advantage of the replicas position information. In \cite{Gamb_Schlegel2013} it has been shown that joint decoding on the collided packets can be attempted, resorting to \ac{MUD} techniques. The authors of \cite{Stefanovic2012,Stefanovic2013TCOM} elaborate the concept of \emph{frameless} slotted scheme, i.e. the duration of a frame is not a-priori fixed but the contention ends when the throughput is maximized. Further evolutions of \ac{RA} include the extension to multiple receiver scenarios \cite{Jakovetic2015} and to all-to-all broadcast transmission \cite{Ivanonv2015}. Identification of replicas for slot synchronous \ac{RA} schemes has been addressed in the works \cite{Bui2015,Zidane2016}, where a simple autocorrelation method has been adopted for identifying replicas of the same user.

It was recently observed that time synchronicity can be abandoned while keeping similar protocol operations. A first attempt in this direction has been done with the \ac{CRA} protocol \cite{Kissling2011a}. Transmitters send replicas of their packets with arbitrary delays within a window of fixed duration. Every time the receiver is successful in decoding a packet, the packet is re-encoded, re-modulated and removed from all positions in the received signal, exploiting the replicas position information stored in the header. Interference cancellation possibly allows further packets to be decoded. A similar scheme to \ac{CRA} is proposed by the authors in \cite{De_Gaudenzi_2014_ACRDA}, where the \ac{VF} concept is introduced, i.e. the interval of time in which all user's replicas are sent. Users are synchronized to their local \ac{VF} and are allowed to send their replicas only in discrete positions within the \ac{VF}. The \ac{VF}s of different users are asynchronous.

An evolution of \ac{CRA} called \ac{ECRA} has been presented first in \cite{Clazzer2012}. At the receiver, after \ac{SIC} is carried out in a similar way as in \ac{CRA}, the received signal samples associated with the replicas that cannot be decoded are combined and a new decoding attempt is performed. Selection combining (SC), \ac{EGC} or \ac{MRC} can be used as combining technique, leading to remarkable gains. The main drawback of \ac{ECRA} is the requirement of perfect knowledge of the replicas position prior to decoding.

In this paper we propose a solution to the problem of localizing the replicas position that does not need any reserved field in the header.

The rest of the paper is organized as follows. Section~\ref{sec:ECRA} reviews the \ac{ECRA} protocol and its features. In Section~\ref{sec:sys_mod} we present the system model, the two phase detection technique and we derive the interference-aware soft-correlation detection rule. Section~\ref{sec:num_res} investigates \ac{ECRA} adopting the derived detection techniques via Monte Carlo simulations and is followed by the conclusions in Section~\ref{sec:end}. 
\section{Enhanced Contention Resolution ALOHA}
\label{sec:ECRA}

\begin{figure*}
    \centering
    \includegraphics[width=0.9\textwidth]{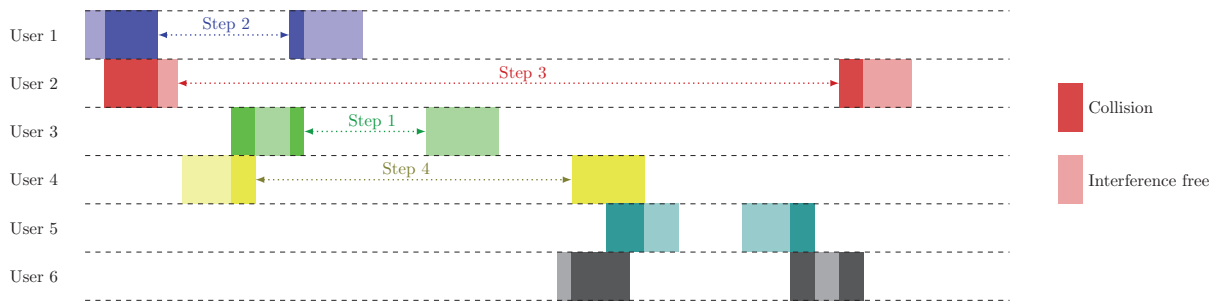}
    \caption{Example of collision pattern at the receiver in \ac{ECRA}, and of the corresponding \ac{SIC} steps. During the first step, user $3$ second replica - the only one free from interference - is decoded and the information content as well as the pointer to its replica are retrieved. Users $1$ and $4$ replicas are now free from interference. In the second step second replica of user $1$ can be decoded. Its interference contribution together with the one of its twin can be now removed from the received signal and first replica of user $2$ is now freed from interference. In step $3$, user $2$ replicas can be decoded and removed from the received signal. Finally in step $4$ user $4$ replicas are decoded and removed from the received signal.}
    \label{fig:MAC_frame}
\end{figure*}

For the sake of simplicity, in the following we assume each user attempting the transmission of one packet only. At the transmitter side each user sends $2$ (or more) replicas of its packet within its local \ac{VF} of duration $T_f$ seconds\footnote{The concept of \ac{VF} has been first introduced in \cite{De_Gaudenzi_2014_ACRDA} and was not present in the first statement of the \ac{ECRA} protocol \cite{Clazzer2012}.} with the \ac{VF} start time only known at the transmitter. The delay between replicas of the same user is chosen at random. An example of a possible received \ac{MAC} signal is shown in Figure \ref{fig:MAC_frame}.

At the receiver side, the \ac{SIC} procedure starts looking for packets that can be decoded. In the example, the first to be found is second replica of user $3$. Once correctly decoded, all replica positions of the decoded user are retrieved from the pointer field in the header. The packet is re-encoded, re-modulated and his waveform is removed from all the identified positions. In this way, the interference caused on the second replica of user $1$ is removed and this replica can be successfully decoded. Similarly, we are able to decode the packets of users $1$ to $4$, while users $5$ and $6$ cannot be decoded.

\begin{figure}
    \centering
    \includegraphics[width=0.4\textwidth]{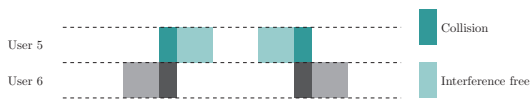}
    \caption{Residual collision pattern after \ac{SIC} decoding for the example in \mbox{Figure \ref{fig:MAC_frame}}.}
    \label{fig:MAC_frame_loop}
\end{figure}

User $5$ and $6$ have both their replicas colliding with each other and \ac{SIC} alone cannot resolve the collision, as emphasized in Figure~\ref{fig:MAC_frame_loop}. With \acs{SC} combining, the interference-free samples from the replicas of user $6$ are selected, creating an enhanced observation of user $6$ packet. On it, decoding is attempted and if successful, its interference contribution is removed from the received signal. This would allow the recovery of user $5$ too.
Other combining techniques, such \ac{MRC} and \ac{EGC} can also be applied, leading to remarkable gains in terms of throughput \cite{Clazzer_ECRA_J2016}. The main challenge for applying combining techniques is the need of performing the detection of the replicas and the identification of the user which they belong to, prior to decoding. A possible way is to exploit the pointer field of the replicas \cite{Clazzer2013}, i.e. by
\begin{itemize}
\item Duplicating the pointer field in the header and trailer of each replica;
\item Protecting the pointer field with a specific low rate \ac{FEC} code.
\end{itemize}
Although viable, both options imply an increase in the protocol overhead, which is critical in applications where the message length is short. We overcome this issue proposing a novel approach that allows detection and localization of replicas without the need of pointer field.
\section{System Model}
\label{sec:sys_mod}


Each user arranges its transmission within a \ac{VF} of duration $T_f$ seconds where the \ac{VF} are asynchronous among users. Each \ac{VF} is divided in $N_s$ slots of duration $\Delta T$, so that $T_f=N_s\Delta T$. Users transmit $d$ replicas of duration $T_p$ seconds within the \ac{VF}. Each replica is transmitted over $n_p$ consecutive slots within the \ac{VF} and we have that a replica duration is a multiple of the slot duration, $T_p=n_p\Delta T$. Each replica is composed by $n_s$ modulated symbols and the symbol duration is $T_s$, so $nT_s=n_p\Delta T=T_p$. Each replica is transmitted starting from a slot index chosen uniformly at random in $[0,N_s-n_p-1]$, rejecting starting slot indexes which lead to self-interference among replicas of a user's packet.

An infinite user population generates traffic following a Poisson process of intensity $\mathsf{G}$. The channel load\footnote{The channel load $\mathsf{G}$ takes into consideration the net information transmitted, depurated from the number of replicas per user $d$.} $\mathsf{G}$ is measured in packet arrivals per packet duration or per $T_p$ seconds.
\begin{figure*}
    \centering
    \includegraphics[width=0.9\textwidth]{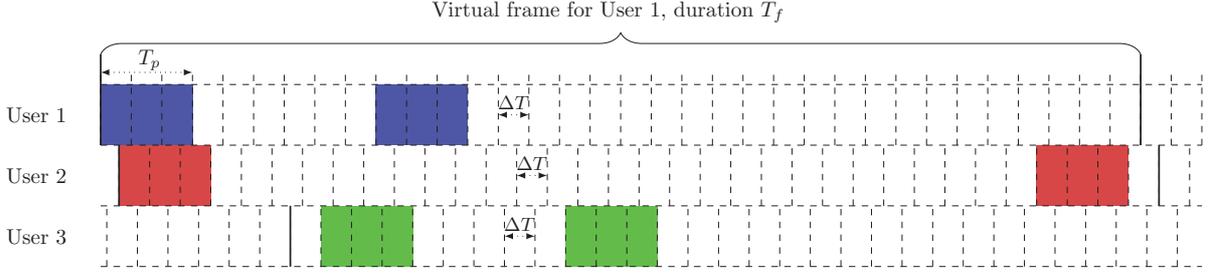}
    \caption{Transmitted signals. Each user sends two replicas of duration $T_p$ seconds that occupy $3$ time slots in the example.}
    \label{fig:MAC_frame_tx}
\end{figure*}
In contrast to \ac{CRA} \cite{Kissling2011a} and the first version of \ac{ECRA} \cite{Clazzer2012}, no pointer field is required in the header for localizing the replicas position. The first section of each replica is a sync word composed by $n_{\sw}$ binary symbols $\mathbf{s}=(s_0,...,s_{n_{\sw}-1})$ common to all users, with $s_i\in \{-1,+1\}$ for $i=0,...,n_{\sw}-1$. The sync word is then appended to the BPSK modulated data part and sent through an \ac{AWGN} channel. The data part carries the actual information and the redundancy introduced by a \ac{FEC} code. The target application of this work is satellite communication links where \ac{AWGN} is a typical channel model extensively used.

Perfect power control is assumed, so that all packets are received with the same power. For a generic user's signal both frequency offset $f$ and epoch $\epsilon$ are modeled as uniform random variables with $f \sim \mathcal{U}\left[-f_{\mathrm{max}};f_{\mathrm{max}}\right]$ and $\epsilon \sim \mathcal{U}\left[0;T_s\right)$. The frequency offset and epoch are common to a user's packet, while they are independent across users. The random phase offset is uniformly distributed between $0$ and $2\pi$, i.e. $\varphi \sim \mathcal{U}\left[0;2\pi\right)$, and it is assumed to be independent replica by replica. Assuming $f_{\mathrm{max}}T_s\ll 1$, the received signal $y(t)$ after matched filtering can be approximated as
\begin{align*}
\label{eq:rx_signal}
y(t)&\cong\sum_{u=1}^{m} \sum_{r=1}^{d} x^{(u)}(t-\epsilon^{(u)}-T^{(u,r)}-t_0^{(u)})e^{j(2\pi f^{(u)}t + \varphi^{(u,r)})}\\
 &+ n(t)
\end{align*}
where,
\begin{itemize}
\item $\epsilon^{(u)}$ is the user $u$ epoch;
\item $T^{(u,r)}$ is the $u$-th user $r$-th replica delay w.r.t. the associated \ac{VF} start;
\item $t_0^{(u)}$ is the user $u$ delay w.r.t. the common reference time;
\item $f^{(u)}$ is the frequency offset for user $u$;
\item $\varphi^{(u,r)}$ is the phase offset for the $r$-th replica of user $u$'s packet.
\end{itemize}
The signal for the $u$-th user $x^{(u)}$ is given by
\begin{equation*}
x^{(u)}(t)= \sum_{i=0}^{n_s-1} a_i^{(u)}g(t-iT_s).
\end{equation*}
Here, $\left\{a_{i}^{(u)}\right\}$ is the symbol sequence forming user $u$'s packet and $g(t)=\mathcal{F}^{-1}\left\{\mathtt{CR}(f)\right\}$ is the pulse shape, where $\mathtt{CR}(f)$ is the frequency response of the raised cosine filter. The noise $n(t)$ is given by $n(t)\triangleq \nu(t) \ast h(t)$, being $\nu(t)$ a white Gaussian process with single-sided power spectral density $N_0$ and $h(t)$ the \ac{MF} impulse response, ${h(t)=\mathcal{F}^{-1} \left\{\sqrt{\mathtt{CR}(f)}\right\}}$.

\subsection{Detection and Decoding}
\label{sec:rx}
At the receiver side, the incoming signal $y(t)$ is sampled and input to the frame start detector. The receiver will operate with a sliding window, similarly to \cite{Meloni2012, De_Gaudenzi_2014_ACRDA}. The decoder starts operating on the first $W$ samples, with $W$ the designed window size. First it detects candidate replicas.
\begin{figure*}
\centering
 \subfigure[Non-coherent soft-correlator used for the detection of candidates replicas.]
   {\includegraphics[width=0.9\textwidth]{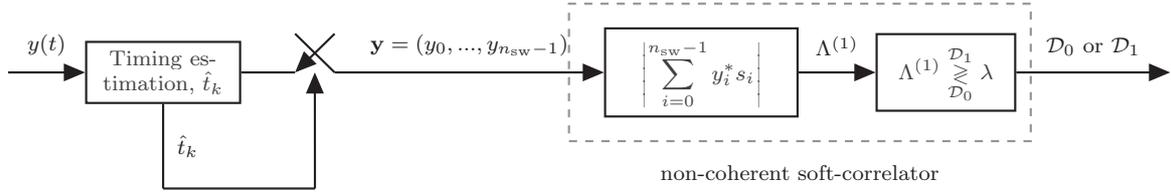}
   \label{fig:rule1}}
 \subfigure[Interference-aware soft-correlator used for the detection of candidates replicas.]
   {\includegraphics[width=0.9\textwidth]{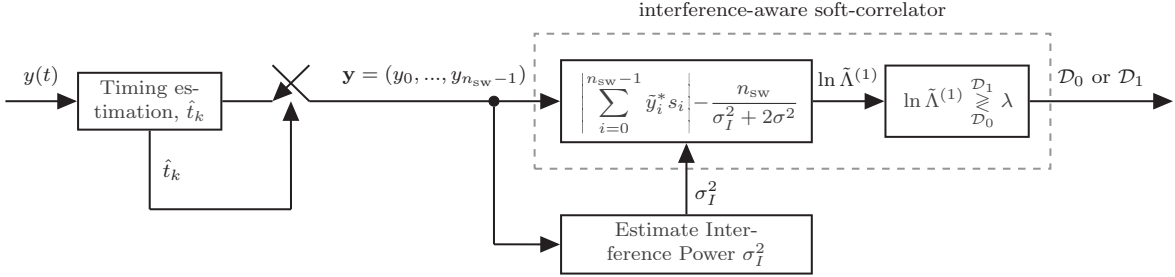}
   \label{fig:rule2}}
 \caption{The non-coherent soft-correlator and the interference-aware soft-correlator used for the detection of candidate replicas.}
 \label{fig:rules}
\end{figure*}

\subsubsection{Detection Phase}
In the first phase the non-coherent soft-correlation metric \cite{Chiani_2010} is used for identifying candidates replicas (see Figure \ref{fig:rule1}). Within a receiver window, a threshold-based test is applied to each of the $W-n_{\sw}$ sequences of $n_{\sw}$ consecutive samples (referred in the following as test intervals) to detect the presence of a sync word. We denote with
\begin{equation*}
\mathbf{y}=(y_0,...,y_{n_{\sw}-1})
\end{equation*}
the sequence of $n_{\sw}$ samples on which the threshold test is applied. Here, we are implicitly assuming that the epoch is estimated prior to frame synchronization. Under the hypothesis that the test interval is aligned to a sync word, the epoch estimation can be reliably performed using pilot-aided\footnote{Observe that the sync word can be effectively used as pilot field for timing estimation.} techniques mutated from code synchronization algorithms used in spread-spectrum communications, see e.g. \cite{Polydoros1984} and references therein. If the test window is not aligned with the sync word of any user, we assume the epoch estimator returning a random sampling offset, uniformly-distributed in $(0,T_s]$. For each test interval - similarly to \cite{Chiani_2010} - the frame synchronizer has to decide among two hypothesis, i.e.
\[
\begin{aligned}
\mathcal{H}_0&:\mathbf{y}=\mathbf{z}+\mathbf{n}\\
\mathcal{H}_1&:\mathbf{y}=\mathbf{s}\,e^{j\varphi^{(u,r)}}+\mathbf{z}+\mathbf{n}
\end{aligned}
\]
where the first hypothesis refers to the case of no sync word, while the second one refers to the case of sync word present. Here $\mathbf{n}=(n_0,...,n_{\sw-1})$ are samples of a discrete white Gaussian process with $n_i\sim \mathcal{CN}(0,2\sigma^2)$ and $\mathbf{z}$ is the interference contribution over the $n_{\sw}$ observed samples.

We adopt the threshold test
\begin{equation}
\label{eq:corr}
\Lambda^{(1)}(\mathbf{y})=\left|\sum_{i=0}^{n_{\sw}-1}y_{i}^* s_i\right|\underset{\mathcal{D}_0}{\overset{\mathcal{D}_1}{\gtrless}}\lambda.
\end{equation}
Where decision $\mathcal{D}_1$ corresponds to hypothesis $\mathcal{H}_1$ and decision $\mathcal{D}_0$ corresponds to hypothesis $\mathcal{H}_0$ and the threshold $\lambda$ is the discriminant between the two decision regions. We call ${\mathcal{S}=\left\{\tau_1 ,\tau_2,...\right\}}$ the set of candidate replica starting positions, i.e. the set containing the positions within the receiver window for which the test of eq.\eqref{eq:corr} outputs $\mathcal{D}_1$. The set of candidate replica positions is the outcome of the first phase.

\subsubsection{Replica Matching Phase}
Let us consider the first candidate replica identified in the first phase. We denote its starting position as $\tau_1$, with $\tau_1 \in \mathcal{S}$. The focus is in finding a subset $\mathcal{S}_1\subseteq \mathcal{S}$ containing the initial positions of bursts that are likely replicas of the (hypothetical) burst starting in position $\tau_1$. To do so, we define the following compatibility criterion:
\begin{definition}[Compatibility Criterion] \label{def:comp_crit}
	A start position $\tau_i \in \mathcal{S}$ is said to be compatible with $\tau_1$ iff
\begin{equation}
\tau_i=\tau_1+k \Delta T
\end{equation}
for some positive integer $k$, $\tau_1<\tau_i<W T_s-\Delta T$.
\end{definition}
The set $\mathcal{S}_1$ is hence formally defined as
\begin{equation}
\mathcal{S}_1\triangleq \left\{\tau_i \in \mathcal{S}| \tau_i=\tau_1+k \Delta T, k \in \mathbb{Z}^+\right\}.
\end{equation}
The subset $\mathcal{S}_1$ contains the starting positions that are compatible (given the \ac{VF} structure) with $\tau_1$, i.e., their associated burst are likely replicas of the burst starting at position $\tau_1$.

Denote with $\mathbf{y}^{(i)}=(y_0^{(i)},...,y_{n_s-1}^{(i)})$ the $n_s$ samples of the received signal starting in position $\tau_i$ within the window. For each $\tau_i \in \mathcal{S}_1$, we compute the non-coherent correlation
\begin{equation}
\label{eq:corr_comb}
\Lambda^{(2)}_{1,i}(\mathbf{y})\triangleq\left|\sum_{j=0}^{n_{s}-1}y_{j}^{(1)}\left[y_j^{(i)} \right]^*\right|.
\end{equation}
We order the $\Lambda^{(2)}_{1,i}$ in descending order and we mark the first $d-1$ as replicas of the same user.

On these replicas we apply combining techniques as \acs{SC}, \ac{MRC} or \ac{EGC}. If decoding is successful, all the replicas are removed from the received signal. Accordingly, $\mathcal{S}$ is updated by removing the starting positions of the cancelled replicas. The process is iterated until $\mathcal{S}$ is empty, or if decoding fails for all remaining candidates in $\mathcal{S}$. The channel decoder is assumed to be capable of identifying unrecoverable errors with high probability.\footnote{Error detection can be implemented either by using an incomplete channel decoder or by concatenating an outer error detection code with the inner channel code.} Once no more packets can be decoded within the window, the receiver window is shifted forward by $\Delta W$ samples and the procedure starts again.

\subsection{Hypothesis Testing, Interference-Aware Rule}
\label{sec:hyp}
We derive here an advanced correlation rule, named $\tilde{\Lambda}^{(1)}$, which takes into consideration the presence of interference. We resort to a Gaussian approximation of the interference contribution. The interference term $\nu_i$ is modeled as $\nu_i\sim \mathcal{CN}(0,\sigma_I^2)$. Furthermore, we assume $\sigma_I^2$ to be constant for the entire test interval. The joint noise and interference contribution is given by $n'_i=\nu_i+n_i$, so that $n'_i\sim \mathcal{CN}(0,\sigma_I^2+2\sigma^2)$.
The approximate \ac{LRT} is then obtained by evaluating,
\begin{equation}
\label{eq:adv_corr1}
\tilde{\Lambda}^{(1)}(\mathbf{y})= \frac{f_{\mathbf{Y}|\mathcal{H}_1}(\mathbf{y}|\mathcal{H}_1)} {f_{\mathbf{Y}|\mathcal{H}_0}(\mathbf{y}|\mathcal{H}_0)} \underset{\mathcal{D}_0}{\overset{\mathcal{D}_1}{\gtrless}} \lambda'
\end{equation}
where $f_{\mathbf{Y}|\mathcal{H}_i}(\mathbf{y}|\mathcal{H}_i)$ is the approximate distribution of the random vector $\mathbf{Y}=(Y_0,...,Y_{n_{\sw}-1})$ under the hypothesis $\mathcal{H}_i$. For the $\mathcal{H}_0$ hypothesis we can write
\begin{equation}
\label{eq:H0}
f_{\mathbf{Y}|\mathcal{H}_0}\left(\mathbf{y}|\mathcal{H}_0\right) = \prod_{i=0}^{n_{\sw}-1}\frac{1}{\pi \left(\sigma_I^2+2\sigma^2\right)} e^{-\frac{|y_i|^2} {\sigma_I^2+2\sigma^2}}.
\end{equation}
For the $\mathcal{H}_1$ hypothesis we can write
\begin{equation}
\label{eq:H1}
f_{\mathbf{Y}|\mathcal{H}_1,\varphi}(\mathbf{y}|\mathcal{H}_1,\varphi) = \prod_{i=0}^{n_{\sw}-1}\frac{1}{\pi \left(\sigma_I^2+2\sigma^2\right)} e^{-\frac{|y_i-s_ie^{j\varphi}|^2} {\sigma_I^2+2\sigma^2}}
\end{equation}
We define $\tilde{y}_i=y_i/\left(\sigma_I^2+2\sigma^2\right)$. Averaging \eqref{eq:H1} over $\varphi$ we find,
\begin{equation}
\begin{split}
\label{eq:H1_2}
f_{\mathbf{Y}|\mathcal{H}_1}(\mathbf{y}|\mathcal{H}_1) &= \left[\prod_{i=0}^{n_{\sw}-1}\frac{1}{\pi \left(\sigma_I^2+2\sigma^2\right)} e^{-\frac{|y_i|^2+1} {\sigma_I^2+2\sigma^2}}\right]\\ &\cdot I_0\left(\left|\sum_{i=0}^{n_{\sw}-1}\tilde{y}_i^*s_i\right|\right).
\end{split}
\end{equation}
Substituting equations \eqref{eq:H1_2} and \eqref{eq:H0} in the expression of equation \eqref{eq:adv_corr1} we get
\begin{equation}
\label{eq:adv_corr2}
\tilde{\Lambda}^{(1)}(\mathbf{y}) = e^{-\frac{n_{\sw}}{\sigma_I^2+2\sigma^2}} I_0\left(\left|\sum_{i=0}^{n_{\sw}-1}\tilde{y}_i^*s_i\right|\right) \underset{\mathcal{D}_0}{\overset{\mathcal{D}_1}{\gtrless}} \lambda'
\end{equation}
Taking the natural logarithm of both sides and making the use of the approximation $\ln (I_0(x))\cong |x|-\ln\sqrt{2\pi |x|}\cong |x|$ \cite{Chiani_2010}, we can rework equation \eqref{eq:adv_corr2} as
\begin{equation}
\label{eq:adv_corr3}
\ln \tilde{\Lambda}^{(1)}(\mathbf{y}) \cong \left|\sum_{i=0}^{n_{\sw}-1}\tilde{y}_i^*s_i\right| - \frac{n_{\sw}}{\sigma_I^2+2\sigma^2} \underset{\mathcal{D}_0}{\overset{\mathcal{D}_1}{\gtrless}} \lambda,
\end{equation}
where $\lambda=\ln\left(\lambda'\right)$. With respect to the non-coherent soft-correlation rule of equation \eqref{eq:corr}, we can observe that in \eqref{eq:adv_corr3} the correlation term is followed by a correction term that depends on the sync word length and on the interference level. The latter is required to be estimated (See Figure \ref{fig:rule2}). 
\section{Numerical Results}
\label{sec:num_res}

We first compare the two non-coherent soft-correlation rules derived in Sections \ref{sec:rx} and \ref{sec:hyp} in terms of \ac{ROC}. In the second part the performance of the \ac{ECRA} receiver in terms of probability of correct detection of the replicas and probability of correct combining of replicas from the same user is shown.

\subsection{\ac{ROC} Comparison}
The performance of the two correlation rules $\Lambda^{(1)}$ and $\tilde{\Lambda}^{(1)}$ that can be adopted in the detection phase of the receiver operations are compared via Monte Carlo simulations. The comparison is done in terms of \ac{ROC}. The false alarm probability $P_F$ is defined as $P_F=\Pr\{\Lambda>\lambda|\mathcal{H}_0\}$. The detection probability $P_D$ is defined as $P_D=\Pr\{\Lambda>\lambda|\mathcal{H}_1\}$. We set $f_{\mathrm{max}} = 0.01/ T_s$. The aggregate signal is then summed with Gaussian noise. The selected $E_s/N_0$ is $E_s/N_0=10$~dB. A sync word of $32$ bits of hexadecimal representation $\{1ACFFC1D\}$ has been adopted, which results in $n_{\sw}=32$ symbols.

\begin{figure}
\centering
 \subfigure[\ac{ROC} for $\Lambda^{(1)}$, $\tilde{\Lambda}^{(1)}$, $\mathsf{G}=0.5$.]
   {\includegraphics[width=0.22\textwidth]{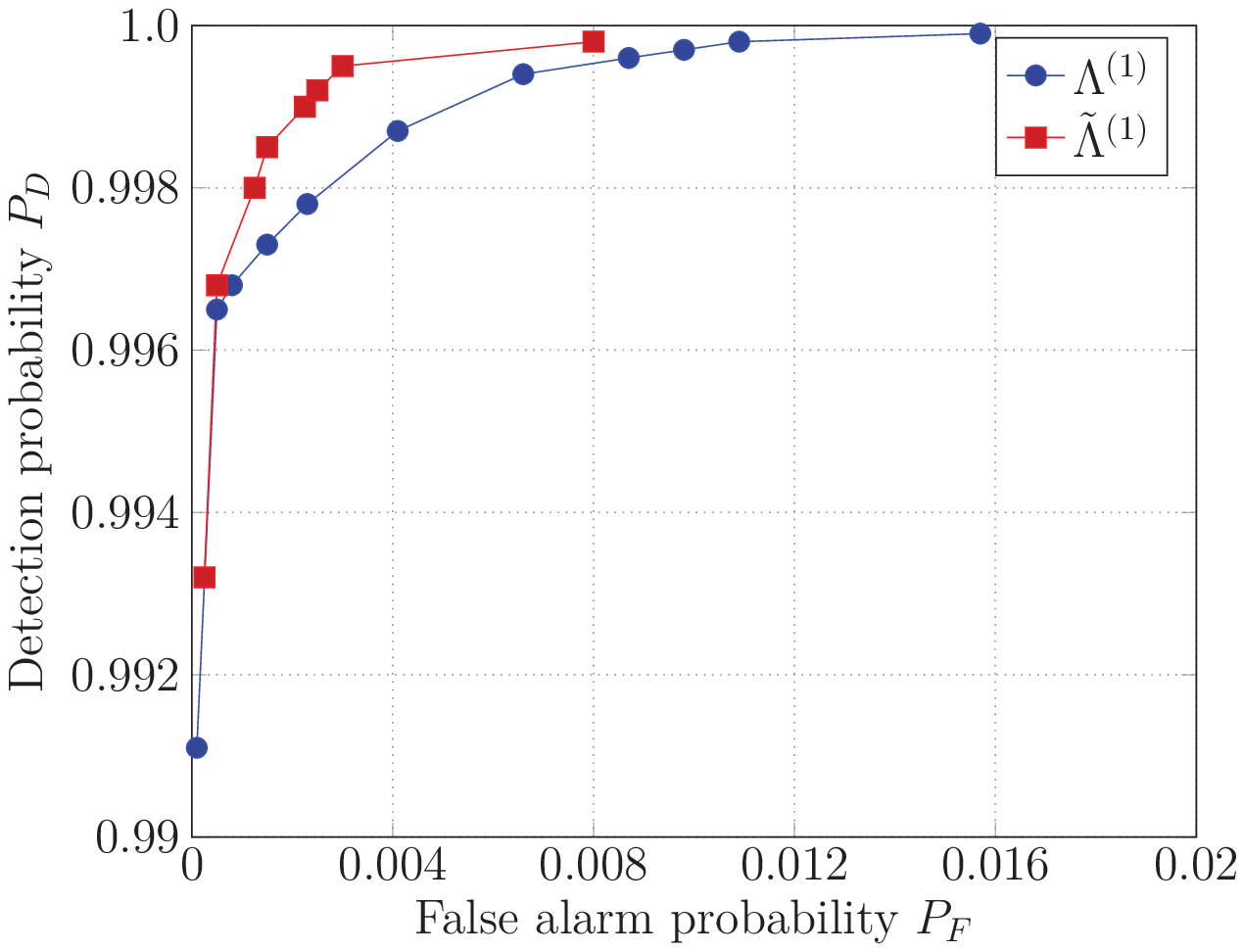}
   \label{fig:Roc_0_5}}
 \subfigure[\ac{ROC} for $\Lambda^{(1)}$, $\tilde{\Lambda}^{(1)}$, $\mathsf{G}=1.5$.]
   {\includegraphics[width=0.22\textwidth]{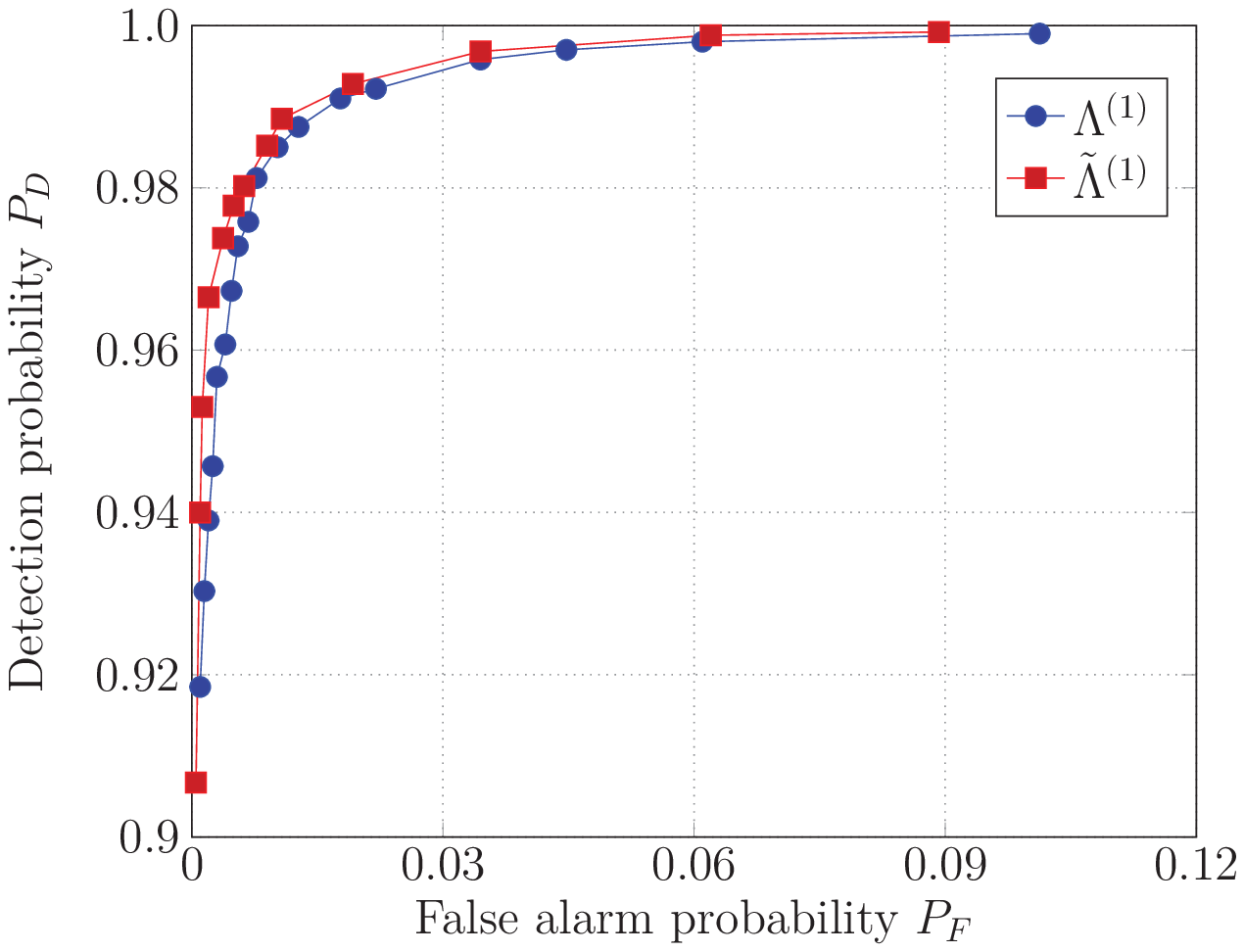}
   \label{fig:Roc_1_5}}
 \caption{\ac{ROC} for the non-coherent and interference-aware soft-correlation synchronization rules, with $\mathsf{G}=\{0.5,1.5\}$, equal received power, $E_s/N_0=10$~dB and $n_{\sw}=32$ symbols.}
 \label{fig:Roc}
\end{figure}

Results for channel traffic values $\mathsf{G}=\{0.5,1.5\}$ are presented in Figure \ref{fig:Roc}. As expected, the knowledge on the interference level exploited in the rule $\tilde{\Lambda}^{(1)}$ leads to better \ac{ROC} performance, regardless form the channel traffic conditions. Nevertheless, the gain compared to the non-coherent correlation rule $\Lambda^{(1)}$ is rather limited. In general, both rules show good performance, with $P_D>0.99$ for $P_F>0.02$ in the worst case (channel traffic $\mathsf{G}=1.5$) and for $\Lambda^{(1)}$.

\subsection{\ac{ECRA} Detection and Replicas Coupling Performance}
\begin{figure}
    \centering
    \includegraphics[width=0.4\textwidth]{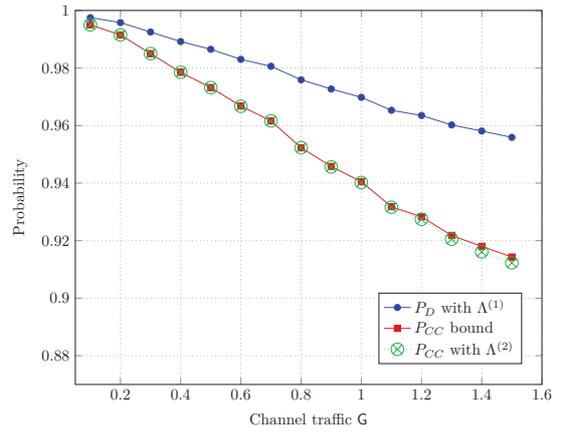}
    \caption{Detection probability $P_D$ for a fixed threshold $\lambda^*$ independent from the channel traffic using $\Lambda^{(1)}$ and correct combining probability $P_{CC}$ with $\Lambda^{(2)}$.}
    \label{fig:Coup_P}
\end{figure}

We present here the results for the detection and correct combining probabilities. We focus in the particular setting where $d=2$ (i.e., users transmit $2$ replicas of their packets). The detection probability $P_D$ has been defined in the previous subsection. We define the correct combining probability $P_{CC}$ as the probability that two replicas of a burst are correctly selected for combining after the two-phase procedure. Obviously, $P_{CC}\leq P_D^2$, i.e., a necessary condition for correct combination is the actual detection of the sync words associated with the two replicas, during the first phase. We select a fixed threshold $\lambda^*$ equal for all the channel traffic values and we use the non-coherent soft-correlation rule $\Lambda^{(1)}$. The threshold $\lambda^*$ has been selected through numerical simulations. We show the results in Figure \ref{fig:Coup_P}, for a \ac{SNR} of $E_s/N_0=10$~dB. The discretization interval equals to one physical layer packet duration, i.e. $\Delta T = T_p$. Each packet is composed by a sync word of $n_{\sw}=32$ symbols (as the one already presented) and a total of $n_s=1000$ BPSK antipodal modulated symbols (including the sync word symbols), the \ac{VF} duration as well as the window duration $W T_s$ are $100$ times the packet duration, $T_f=W T_s=100 T_p$.

Observe that the detection probability remains above $95\%$ for all the channel traffic $\mathsf{G}$ values, up to $\mathsf{G}=1.5$. The non-coherent soft-correlation rule $\Lambda^{(1)}$ is particularly robust to variations in the channel traffic, since the presented results are obtained for a single threshold value $\lambda^*$ which has been kept constant for all the channel traffic values. For all values of channel traffic simulated, the correct combining probability is very close to the bound $P_D^2$.

\subsection{Spectral Efficiency}
We compare the simulation results in terms of both spectral efficiency achieved by \ac{ECRA} with \ac{MRC}, after the two-phase detection process described in Section \ref{sec:rx}. The proposed technique is compared against the idealized case in which all replicas positions are known to the receiver prior to decoding. We select $\Delta T=T_p$ and again the window duration is $T_f=W T_s=100 T_p$. Perfect \ac{CSI} at the receiver is assumed for enabling \ac{MRC}.

We adopt the non-coherent soft-correlation rule $\Lambda^{(1)}$ and a fixed threshold kept constant, regardless the channel load $\mathsf{G}$. All replicas are received with equal power $E_S/N_0=2$ dB. A capacity achieving code adopting a Gaussian codebook with rate $R=1$ is assumed, so that if the mutual information at the output of the combiner exceeds the rate $R$, then the packet is considered to be successfully decoded. Further refinements of the decoding model can be adopted following a realistic \ac{PLR} performance of a specific code for example. Nonetheless, for the present work such a model is sufficient to show the goodness of the detection and identification approach. The maximum number of \ac{SIC} iterations is set to $10$. \ac{SIC} is assumed ideal. That is, if the position of both replicas of one user is known at the receiver, \ac{MRC} is applied and if the packet can be decoded its interference contribution is fully removed from the received signal.
\begin{figure}
    \centering
    \includegraphics[width=0.4\textwidth]{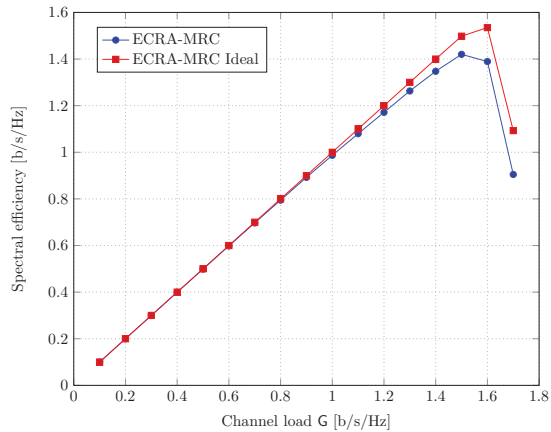}
    \caption{Spectral efficiency of \ac{ECRA}-\ac{MRC} with the proposed two phase detection and combining technique compared to the ideal \ac{ECRA}-\ac{MRC}.}
    \label{fig:Thr}
\end{figure}

In Figure \ref{fig:Thr}, the spectral efficiency results for the proposed two phase detection and combining technique (called \ac{ECRA}-\ac{MRC} in the legend) and the ideal \ac{ECRA}-\ac{MRC} where all the replica positions is known at the receiver are presented. The proposed technique is close to the performance of the ideal case. The maximum spectral efficiency exceeds $1.4$ b/s/Hz, which is only $8\%$ less than the maximum spectral efficiency of the ideal case. 
\section{Conclusion}
\label{sec:end}

A solution for localizing candidate replicas and combine them prior to decoding is presented. In the asynchronous random access protocol \ac{ECRA}, it allows the exploitation of combining techniques as \ac{MRC}. A two phase approach is proposed. First candidate replicas are identified using the known sync word. Non-coherent soft-correlation is adopted as baseline metric and an interference-aware soft-correlation rule is derived. The latter can be adopted when the interference power on the sync word can be estimated. Numerical results have shown that already the simple non-coherent soft-correlation metric is sufficient to guarantee the detection of most replicas. For example,  up to $99.5\%$ of replicas can be detected for a channel load of $\mathsf{G}=1$ b/s/Hz. In the second phase, the entire candidate replica signal is exploited to compute the non-coherent soft-correlation metric against the other candidates. 




\bibliographystyle{./bibtex/IEEEtran}
\bibliography{IEEEabrv,./References}

\begin{thebibliography}{10}
\providecommand{\url}[1]{#1}
\csname url@samestyle\endcsname
\providecommand{\newblock}{\relax}
\providecommand{\bibinfo}[2]{#2}
\providecommand{\BIBentrySTDinterwordspacing}{\spaceskip=0pt\relax}
\providecommand{\BIBentryALTinterwordstretchfactor}{4}
\providecommand{\BIBentryALTinterwordspacing}{\spaceskip=\fontdimen2\font plus
\BIBentryALTinterwordstretchfactor\fontdimen3\font minus
  \fontdimen4\font\relax}
\providecommand{\BIBforeignlanguage}[2]{{%
\expandafter\ifx\csname l@#1\endcsname\relax
\typeout{** WARNING: IEEEtran.bst: No hyphenation pattern has been}%
\typeout{** loaded for the language `#1'. Using the pattern for}%
\typeout{** the default language instead.}%
\else
\language=\csname l@#1\endcsname
\fi
#2}}
\providecommand{\BIBdecl}{\relax}
\BIBdecl

\bibitem{Abramson1970}
N.~Abramson, ``{The ALOHA system: Another alternative for computer
  communications},'' in \emph{Proceedings of the 1970 Fall Joint Comput. Conf.,
  AFIPS Conf.}, vol.~37, Montvale, N.~J., November 1970, pp. 281--285.

\bibitem{Roberts1975}
L.~G. Roberts, ``{ALOHA packet system with and without slots and capture},''
  \emph{SIGCOMM Comput. Commun. Rev.}, vol.~5, pp. 28--42, April 1975.

\bibitem{Casini2007}
E.~Casini, R.~De~Gaudenzi, and O.~del Rio~Herrero, ``{Contention Resolution
  Diversity Slotted {ALOHA} ({CRDSA}): An Enhanced Random Access Scheme for
  Satellite Access Packet Networks},'' \emph{{IEEE Transactions on Wireless
  Communications}}, vol.~6, no.~4, pp. 1408--1419, April 2007.

\bibitem{Liva2011}
G.~Liva, ``{Graph-Based Analysis and Optimization of Contention Resolution
  Diversity Slotted ALOHA},'' \emph{IEEE Transactions on Communications},
  vol.~59, no.~2, pp. 477--487, February 2011.

\bibitem{Paolini_2015_TIT}
E.~Paolini, G.~Liva, and M.~Chiani, ``{Coded Slotted ALOHA: A Graph-Based
  Method for Uncoordinated Mutliple Access},'' \emph{IEEE Transactions on
  Information Theory}, vol.~61, no.~12, pp. 6815--6832, October 2015.

\bibitem{Narayanan_2012}
K.~Narayanan and H.~Pfister, ``{Iterative Collision Resolution for Slotted
  ALOHA: An Optimal Uncoordinated Transmission Policy},'' in \emph{7th
  International Symposium on Turbo Codes and Iterative Information Processing
  (ISTC)}, Gothenburg, Sweden, August 2012, pp. 136--139.

\bibitem{Gamb_Schlegel2013}
M.~Ghanbarinejad and C.~Schlegel, ``{Irregular Repetition Slotted ALOHA with
  Multiuser Detection},'' in \emph{10th Annual Conference on Wireless On-demand
  Network Systems and Services (WONS)}, Banff, AB, March 2013, pp. 201--205.

\bibitem{Stefanovic2012}
C.~Stefanovic, P.~Popovski, and D.~Vukobratovic, ``{Frameless ALOHA Protocol
  for Wireless Networks},'' \emph{IEEE Communication Letters}, vol.~16, no.~12,
  pp. 2087--2090, October 2012.

\bibitem{Stefanovic2013TCOM}
C.~Stefanovic and P.~Popovski, ``{ALOHA Random Access that Operates as a
  Rateless Code},'' \emph{IEEE Transactions on Communications}, vol.~61,
  no.~11, pp. 4653--4662, November 2013.

\bibitem{Jakovetic2015}
D.~Jakovetic, D.~Bajovic, D.~Vukobratovic, and V.~Crnojevic, ``{Cooperative
  Slotted Aloha for Multi-Base Station Systems},'' \emph{IEEE Transactions on
  Communications}, vol.~63, no.~4, pp. 1443--1456, April 2015.

\bibitem{Ivanonv2015}
M.~Ivanov, F.~Br\"annstr\"om, G.~Graell~i Amat, and P.~Popovski, ``{All-to-all
  Broadcast for Vehicular Networks Based on Coded Slotted ALOHA},'' in
  \emph{IEEE ICC Workshop on Massive Uncoordinated Access Protocols (MASSAP)},
  May 2015, pp. 2046--2050.

\bibitem{Bui2015}
H.-C. Bui, K.~Zidane, J.~Lacan, and M.-L. Boucheret, ``{A Multi-Replica
  Decoding Technique for Contention Resolution Diversity Slotted Aloha},'' in
  \emph{82nd IEEE Vehicular Technology Conference (VTC)}, Boston, MA, USA,
  September 2015, pp. 1--5.

\bibitem{Zidane2016}
K.~Zidane, J.~Lacan, M.~Gineste, C.~Bes, A.~Deramecourt, and M.~Dervin,
  ``{Estimation of Timing Offsets and Phase Shifts Between Packet Replicas in
  MARSALA Random Access},'' in \emph{Available at
  https://arxiv.org/ftp/arxiv/papers/1511/1511.05359.pdf}, 2016.

\bibitem{Kissling2011a}
C.~Kissling, ``{Performance Enhancements for Asynchronous Random Access
  Protocols over Satellite},'' in \emph{2011 IEEE International Conference on
  Communications (ICC)}, Kyoto, Japan, June 2011, pp. 1--6.

\bibitem{De_Gaudenzi_2014_ACRDA}
R.~De~Gaudenzi, O.~del Rio~Herrero, G.~Acar, and G.~Barrab\`es, ``{Asynchronous
  Conteniton Resolution Diversity ALOHA: Making CRDSA Truly Asynchronous},''
  \emph{IEEE Transactions on Wireless Communications}, vol.~13, no.~11, pp.
  6193--6206, November 2014.

\bibitem{Clazzer2012}
F.~Clazzer and C.~Kissling, ``{Enhanced Contention Resolution ALOHA} -
  {ECRA},'' in \emph{2013 International ITG Conference on Systems,
  Communications and Coding (SCC)}, Munich, Germany, January 2013, pp. 1--6.

\bibitem{Clazzer_ECRA_J2016}
F.~Clazzer, C.~Kissling, and M.~Marchese, ``{Exploiting Combination Techniques
  in Random Access MAC Protocols: Enhanced Contention Resolution ALOHA},''
  \emph{Available at http://arxiv.org/pdf/1602.07636.pdf}, 2016.

\bibitem{Clazzer2013}
F.~Clazzer and C.~Kissling, ``{Optimum Header Positioning in Successive
  Interference Cancellation (SIC) based ALOHA},'' in \emph{2013 IEEE
  International Conference on Communications (ICC)}, Budapest, Hungary, June
  2013, pp. 2869--2874.

\bibitem{Meloni2012}
A.~Meloni, M.~Murroni, C.~Kissling, and M.~Berioli, ``{Sliding Window-Based
  Contention Resolution Diversity Slotted ALOHA},'' in \emph{Proc. IEEE Global
  Comm. Conf. (GLOBECOM)}, Anaheim, CA, USA, December 2012, pp. 3305--3310.

\bibitem{Chiani_2010}
M.~Chiani, ``{Noncoherent Frame Synchronization},'' \emph{IEEE Transactions on
  Communications}, vol.~58, no.~5, pp. 1536--1545, May 2010.

\bibitem{Polydoros1984}
A.~Polydoros and C.~Weber, ``{A Unified Apporach to Serial Search
  Spread-Spectrum Code Acquisition Part I: General Theory},'' \emph{IEEE
  Transactions on Communications}, vol.~32, no.~5, pp. 542--549, May 1984.

\end{thebibliography}

~


\end{document}